\documentclass[%
 reprint,
amsmath,amssymb,aps,
prb,
]{revtex4-1}

\usepackage{verbatim}
\usepackage{graphicx}
\normalfont \topmargin -0.5 cm
\begin{document}

\title{Dynamic nuclear polarization in InGaAs/GaAs and GaAs/AlGaAs quantum dots under non-resonant ultra-low power optical excitation}

\author{J.~Puebla$^1$}
\author{E.~A.~Chekhovich$^1$}
\email{e.chekhovich@sheffield.ac.uk}
\author{M.~Hopkinson$^2$}
\author{P.~Senellart$^3$, A.~Lemaitre$^3$}
\author{M.~S.~Skolnick$^1$}
\author{A.~I.~Tartakovskii$^1$}
\email{a.tartakovskii@sheffield.ac.uk} \affiliation{$^1$Department
of Physics and Astronomy, University of Sheffield, Sheffield S3
7RH, United Kingdom} \affiliation{$^2$Department of Electronic and
Electrical Engineering, University of Sheffield, Sheffield S1 3JD,
United Kingdom} \affiliation{$^3$Laboratoire de Photonique et
Nanostructures, LPN/CNRS, Route de Nozay, 91460 Marcoussis,
France}
\date{\today}

\begin{abstract}
We study experimentally the dependence of dynamic nuclear spin
polarization on the power of non-resonant optical excitation in
two types of individual neutral semiconductor quantum dots:
InGaAs/GaAs and GaAs/AlGaAs. We show that the mechanism of nuclear
spin pumping via second order recombination of optically forbidden
(''dark'') exciton states recently reported in InP/GaInP quantum
dots [Phys. Rev. B 83, 125318 (2011)] is relevant for material
systems considered in this work. In the InGaAs/GaAs dots this
nuclear spin polarization mechanism is particularly pronounced,
resulting in Overhauser shifts up to $\sim$80~$\mu$eV achieved at
optical excitation power $\sim$1000 times smaller than the power
required to saturate ground state excitons. The Overhauser shifts
observed at low-power optical pumping in the interface GaAs/AlGaAs
dots are generally found to be smaller (up to $\sim$40~$\mu$eV).
Furthermore in GaAs/AlGaAs we observe dot-to-dot variation and
even sign reversal of the Overhauser shift which is attributed to
dark-bright exciton mixing originating from electron-hole exchange
interaction in dots with reduced symmetry. Nuclear spin
polarization degrees reported in this work under ultra-low power
optical pumping are comparable to those achieved by techniques
such as resonant optical pumping or above-gap pumping with high
power circularly polarized light. Dynamic nuclear polarization via
second-order recombination of ''dark'' excitons may become a
useful tool in single quantum dot applications, where manipulation
of the nuclear spin environment or electron spin is required.
\end{abstract}

\maketitle

\section{Introduction}

In all III-V semiconductor materials the atoms of the lattice
possess non-zero nuclear spins and, as a result, have non-zero
nuclear magnetic moments. Nuclear magnetism has almost no effect
in classical optoelectronic semiconductor devices, but becomes an
important factor in nano-scale devices, particularly in
semiconductor quantum dots (QDs) where quantum degrees of freedom
of single charge particles and photons are addressed. The major
effect associated with nuclear magnetism is electron-nuclear
(hyperfine) interaction. In quantum dots nuclear spins are capable
of producing effective (Overhauser) magnetic fields as large as a
few tesla, which has a dramatic effect on the quantum mechanical
evolution of the electron trapped in the dot (see recent reviews
\cite{RMPReview,NatMatReview,BrackerSSTReview,CoishPSSb}). As a
consequence a considerable effort has been put recently into
understanding as well as controlling nuclear magnetism in
semiconductor nanostructures
\cite{Imamoglu,LossNarrowing,MarcusSiGe,KouwenhovenNanowire,OnoNMR}.

One very basic operation required to control nuclear spins is
dynamic nuclear polarization (DNP) - a process that can produce a
controllable non-equilibrium alignment (polarization) of nuclear
spins along a certain direction (usually the direction of external
magnetic field). DNP is based on the hyperfine interaction with
electrons: if electrons are spin-polarized (e.g. by optical
orientation) they can transfer their polarization to nuclei via
flip-flops
\cite{AbrahamBook,GammonPRL,Tartakovskii,Lai,Eble,Braun1,InPX0,Imamoglu,Korenev,CarterDNP}.
The main obstacle in achieving high nuclear spin polarization
degrees is a mismatch in Zeeman splittings of electrons and nuclei
that reduces the DNP efficiency: the electron has much larger
g-factor than the nucleus and thus electron-nuclear flip-flop
requires additional energy to be absorbed or emitted.

Several approaches for enhancing DNP efficiency have been
demonstrated in quantum dots. For example continuous fast
injection and removal of spin polarized electrons from the QD can
increase the number of electron-nuclear flip-flops thus
compensating for the low probability of such process. This can be
achieved using non-resonant pumping with circularly polarized
light of a power sufficiently high to overcome fast depolarization
and optical recombination of electrons \cite{InPX0,Tartakovskii}.
Alternatively resonant optical excitation can be used to induce
second-order (forbidden) processes. The key feature of this
approach is direct compensation of the electron-nuclear Zeeman
energy mismatch by the energy of the absorbed photon, which does
not involve high population probability of the dot
\cite{InPRes,Latta,Hogele,Xu1}.

Here we study experimentally a different mechanism of DNP, where
spin polarization is transferred to nuclei from long-lived
optically-forbidden (''dark'') excitons in neutral QDs
\cite{InPX0,KorenevSelfPol,FinleyX0}. Due to their long lifetime
dark states can be efficiently populated even with non-resonant
low-power optical pumping ($\sim$1000 lower than the power
required to populate optically-allowed states). Furthermore,
second-order recombination with simultaneous nuclear spin-flip can
be a dominant process for dark exciton decay since any competing
radiative recombination is very weak. As a result second-order
recombination of dark excitons becomes an efficient DNP mechanism:
we find that in InGaAs dots it leads to nuclear polarization
characterized by Overhauser shifts of up to $\sim$80~$\mu$eV
(fully polarized nuclei would lead to the Overhauser shift in the
range 120$\div$300$\mu$eV depending on indium concentration
\cite{CoishPSSb,Gueron,Gotschy}). In the studied QDs such an
Overhauser shift corresponds to an effective magnetic field of
$\sim$3~tesla. Combined with our previous work on InP/GaInP QDs
\cite{InPX0} the current results obtained on GaAs/AlGaAs and
InGaAs/GaAs dots allow us to conclude that DNP via dark excitons
is a phenomenon universal for neutral QDs. Due to its undemanding
requirements (only low power non-resonant excitation is needed)
DNP via dark excitons may become a versatile tool for controlling
nuclear magnetic environment in quantum dots used to implement
electron/hole spin qubits
\cite{Latta,Vink,Xu1,Stepanenko,Khaetskii,Merkulov,Ladd,PettaCircuit,MedfordDecoupling,CywinskiDephasing,GreilichSingleMode,Taylor,Gerardot,CarterDNP}.
Moreover due to its universality and high efficiency this DNP
mechanism should also be taken into account in those cases where
DNP needs to be avoided.

The rest of the paper is organized as follows: In Section
\ref{sec:Samples} we describe experimental techniques and quantum
dot samples used. In Sec. \ref{sec:PL} we discuss
photoluminescence (PL) spectra of the studied QDs with particular
focus on PL of dark excitons. The main results on DNP in neutral
QDs are presented in Sec. \ref{sec:DNP} in the following order: In
Sec. \ref{subsec:DNPDet} we describe the techniques used to detect
the Overhauser shift in single QDs using PL spectroscopy. In Sec.
\ref{subsec:DNPInGaAs} we present experimental results on DNP in
InGaAs/GaAs QDs and discuss the mechanism of DNP via second-order
recombination of dark excitons. In Sec. \ref{subsec:DNPGaAs} the
results on DNP in GaAs/AlGaAs dots are shown and the difference of
the DNP processes in InGaAs and GaAs dots is discussed. The paper
is concluded in Sec. \ref{sec:Concl}, with additional experimental
data shown in Appendix \ref{App:PDepAddit}.

\section{\label{sec:Samples}Quantum dot samples and experimental techniques}

The experiments were performed on neutral quantum dots in
InGaAs/GaAs and GaAs/AlGaAs nominally undoped samples. Both
structures were grown by molecular beam epitaxy (MBE). GaAs/AlGaAs
interfacial dots were formed by a monolayer (ML) fluctuations of a
9~ML GaAs quantum well embedded in Al$_{0.3}$Ga$_{0.7}$As barriers
(further details on this sample can be found in Refs.
\cite{GaAsDiffusion,MakhoninPRB,HoleNucIso}). The self-assembled
InGaAs/GaAs dots emitting at $\lambda\sim$914~nm were grown in a
low Q-factor (Q$\sim$250) cavity that enhances the quantum dot
luminescence signal (See details in Ref.
\cite{QNMRArxiv,HoleNucIso}).

The structures were investigated using micro-photoluminescence
($\mu$-PL) spectroscopy of single neutral quantum dots. All
experiments were carried out in a helium gas-exchange cryostat at
$T$=4.2~K. The sample was excited by the laser focused by an
aspheric lens into a spot $\sim$1~$\mu$m in diameter. The same
lens was used to collect the PL signal which was then analyzed by
a double spectrometer with a 1~meter working distance and equipped
with a back-illuminated deep-depletion charge-coupled device (CCD)
camera. The excitation energy was chosen to match wetting layer
for InGaAs dots ($E_{exc}$=1.46~eV) and was above the quantum well
ground state energy ($E_{exc}$=1.73~eV) for GaAs dots. Magnetic
field $B_z$ up to 8~T was applied normal to the sample surface and
parallel to the direction of PL excitation and collection (Faraday
geometry).

Several individual QDs have been examined in both InGaAs (dots are
labeled A1, A2, A3, etc. throughout the text) and GaAs (dots are
labeled B1, B2, etc.) samples to verify the systematic nature of
DNP at ultra-low optical powers. More detailed measurements have
been done for five individual dots from each sample and yielded
qualitatively similar results. Therefore the discussion of the
main text is focused on the data obtained from dots A1, A2 and B1,
while more experimental results obtained on other individual dots
are presented in Appendix \ref{App:PDepAddit}.

\section{\label{sec:PL}Photoluminescence spectroscopy of dark and bright exciton states in neutral quantum dots}

In a neutral quantum dot electrons $\uparrow$($\downarrow$) and
heavy holes $\Uparrow$($\Downarrow$) in a spin up (down) state
along the growth axis $Oz$ form either optically forbidden
''dark''
\cite{Bayer,Poem,InPX0,KorenevSelfPol,HoleNucIso,InPHoleNuc,SmolenskiDark,KurtzeDark,SanchoDark,FinleyX0,Karlsson,Besombes,Witek,BayerDark}
excitons $\left|\Uparrow\uparrow\right>$
($\left|\Downarrow\downarrow\right>$) with spin projection
$+2(-2)$, or ''bright'' excitons $\left|\Uparrow\downarrow\right>$
($\left|\Downarrow\uparrow\right>$) with $+1(-1)$ spin projection.
The structure of the exciton eigenstates in the dot is determined
by the electron-hole ($e-h$) exchange interaction
\cite{GammonPRL,Bayer}. In quantum dots with low symmetry,
exchange interaction mixes bright and dark states
\cite{Bayer,InPDyn,InPX0}. As a result dark excitons gain dipole
oscillator strength (due to admixture of
$\left|\Uparrow\downarrow\right>$ and
$\left|\Downarrow\uparrow\right>$ states) and become visible in
PL.

\begin{figure}
\includegraphics{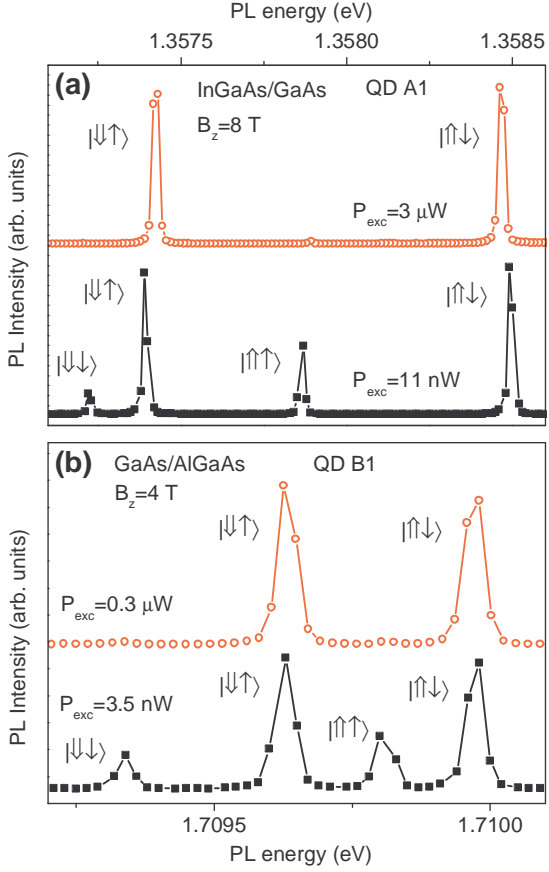}
\caption{\label{fig:Spectra}Photoluminescence (PL) spectra of
neutral InGaAs quantum dot A1 (a) and GaAs QD B1 (b) measured in
external magnetic field applied along the sample growth axis
$B_z$=8~T or $B_z$=4~T, respectively. Two spectra are shown for
each quantum dot: at ultra-low optical excitation power
($P_{exc}$=11~nW for InGaAs and $P_{exc}$=3.5~nW for GaAs) and at
high power ($P_{exc}$=3~$\mu$W for InGaAs and $P_{exc}$=0.3~$\mu$W
for GaAs). Spectra are normalized to have similar maximum
intensities. In a neutral quantum dot heavy hole $\Uparrow$
($\Downarrow$) and electron $\uparrow$($\downarrow$) with spins
parallel (antiparallel) to external field can form optically
allowed ''bright'' states
($|\Uparrow\downarrow\rangle$,$|\Downarrow\uparrow\rangle$) and
forbidden ''dark'' states
($|\Uparrow\uparrow\rangle$,$|\Downarrow\downarrow\rangle$) with
total spin projections $J_{Z}$ = $\pm1$ and $\pm2$ respectively.
At ultra-low excitation powers all four (bright and dark) excitons
are observed in PL. At high powers PL of dark states is saturated
and only bright states are observed.}
\end{figure}

Typical PL spectra of single neutral QDs measured in magnetic
field $B_z$ along the sample growth axis are shown in Fig.
\ref{fig:Spectra} for InGaAs/GaAs (a) and GaAs/AlGaAs (b) samples.
For each QD two spectra are presented: at ultra-low excitation
power ($P_{exc}$=11~nW for InGaAs and $P_{exc}$=3.5~nW for GaAs)
and at high power ($P_{exc}$=3~$\mu$W for InGaAs and
$P_{exc}$=0.3~$\mu$W for GaAs).

We start with discussion of the spectra measured at ultra-low
powers. At these powers QD is in ''linear'' regime, i.e. PL
intensity of all exciton lines depend linearly on excitation power
$P_{exc}$ (Refs. \cite{PowerDep,InPX0}). Such regime is realized
when the total probability to find the dot occupied by an exciton
(in any spin state) is much less then unity ($\ll 1$). Under these
conditions PL intensity of each exciton state will be determined
by two factors (i) the probability for this state to be populated
by the laser excitation and (ii) non-radiative escape rate
(spin-flips, or non-radiative recombination). If the rate of
non-radiative processes is negligibly small the relative PL
intensity of each exciton state will be proportional to its
initial population probability after the optical excitation. This
is because each exciton (dark or bright) will have sufficient time
to emit a photon if the dot is in the linear regime. As a result
at ultra-low powers all four possible excitons have comparable PL
intensities (Fig. \ref{fig:Spectra}). We note that observation of
dark excitons in InGaAs/GaAs and GaAs/AlGaAs QDs at ultra-low
optical powers complement the previous report for InP/GaInP
quantum dots \cite{InPX0}, demonstrating that this phenomenon is
not specific to a certain QD material system.

When optical excitation power is increased the PL of dark excitons
saturates due to their small optical recombination rate. This
saturation takes place when QD is no longer in linear regime, and
dark excitons can be effectively depopulated via capture of a
second exciton and formation of a biexciton state. As a result
dark excitons have relatively small intensity compared to bright
sates at increased optical power (Fig. \ref{fig:Spectra}).

\begin{figure}
\includegraphics{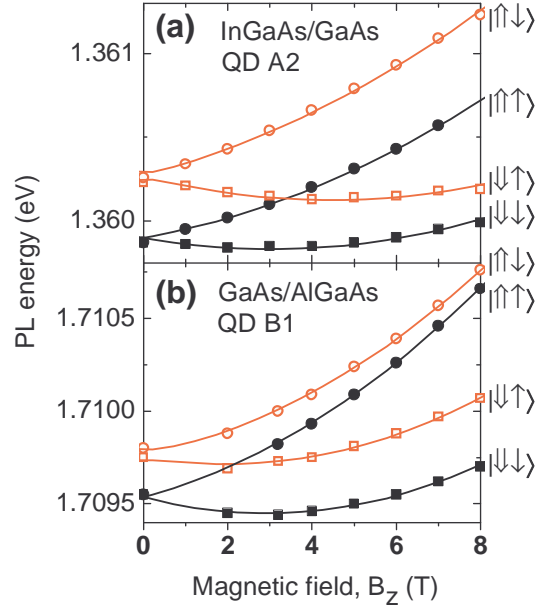}
\caption{\label{fig:BDep}Magnetic field dependence of exciton PL
energies in InGaAs QD A2 (a) and GaAs QD B1 (b). Open symbols
represent bright states $|\Uparrow\downarrow\rangle$ (circles) and
$|\Downarrow\uparrow\rangle$ (squares), while solid symbols
correspond to dark states $|\Uparrow\uparrow\rangle$ (circles) and
$|\Downarrow\downarrow\rangle$ (squares). Lines show fitting using
Eqs. \ref{eq:Zeeman} allowing electron and hole g-factors to be
determined (see details in text).}
\end{figure}

The dependence of PL energies of dark ($E_d$) and bright ($E_b$)
exciton states on external magnetic field $B_z$ is shown with
symbols in Fig. \ref{fig:BDep} for InGaAs/GaAs (a) and GaAs/AlGaAs
(b) samples. We use the following model equations for exciton
energies \cite{Bayer,InPX0,BayerDark}:
\begin{eqnarray}
\label{eq:Zeeman} E_{b}=E_{0}+\kappa
B^2_{z}+\frac{\delta_{0}}{2}\pm\frac{1}{2}\sqrt{\delta^2_{b}+\mu^2_{B}(g_{h}-g_{e})^2B^2_{z}},\nonumber\\
E_{d}=E_{0}+\kappa
B^2_{z}-\frac{\delta_{0}}{2}\pm\frac{1}{2}\sqrt{\delta^2_{d}+\mu^2_{B}(g_{h}+g_{e})^2B^2_{z}},
\end{eqnarray}
where $\mu_{B}$ is Bohr magneton, $E_0$ -- QD band-gap energy,
$\kappa$ -- diamagnetic constant, $g_e$($g_h$) -- electron (hole)
g-factor, $\delta_0$ is the splitting between dark and bright
exciton doublets and $\delta_d$ ($\delta_b$) is the dark (bright)
doublet fine structure splitting. Since the Zeeman splitting of
dark (bright) excitons is determined by the sum (difference) of
$g_e$ and $g_h$ electron and hole g-factors can be determined
independently from the experiment. The results of fitting using
Eqs. \ref{eq:Zeeman} are shown in Fig. \ref{fig:BDep} with lines.
We find the following fitting parameters:
$\kappa\approx7~\mu$eV/T$^2$, $\delta_{0}\approx370~\mu$eV,
$\delta_{b}\approx35~\mu$eV, $g_{e}=-0.35$ and $g_{h}=1.9$ for
InGaAs/GaAs QD A2 and $\kappa\approx10~\mu$eV/T$^2$,
$\delta_{0}\approx230~\mu$eV, $\delta_{b}\approx55~\mu$eV,
$g_{e}=0.3$ and $g_{h}=1.8$ for GaAs/AlGaAs QD B1. The splitting
$\delta_d$ could not be resolved and was fixed to zero during the
fitting. From the measurements on several other dots from the same
samples we find very similar values of $\kappa$, $\delta_{0}$,
$g_{e}$ and $g_{h}$ while fine structure splitting $\delta_{b}$
changes considerably from dot to dot.

\section{\label{sec:DNP}Dynamic nuclear polarization in neutral quantum dots at ultra-low excitation power}
\subsection{\label{subsec:DNPDet}Detection of nuclear spin polarization in quantum dots}

\begin{figure}
\includegraphics{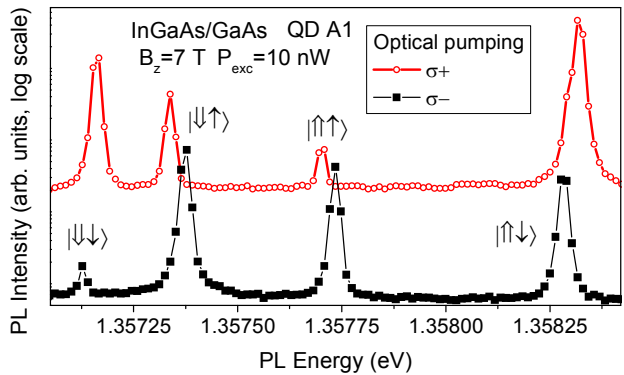}
\caption{\label{fig:BNDet} Detection of nuclear spin polarization
using PL in a single quantum dot. PL spectra of the InGaAs QD A1
measured at magnetic field $B_z=7$~T and excitation power
$P_{exc}$=10~nW are shown for $\sigma^+$ (open symbols) and
$\sigma^-$ (solid symbols) polarized optical pumping. Logarithmic
vertical scale is used to make all dark and bright exciton PL
lines clearly visible in both spectra. At $\sigma^-$ polarized
optical pumping nuclear spin polarization is close to zero, while
$\sigma^+$ pumping induces significant negative nuclear spin
polarization antiparallel to external field (see Sec.
\ref{subsec:DNPInGaAs}). This nuclear spin polarization shifts
excitons with electron spin $\uparrow$ ($\downarrow$) to lower
(higher) energies. We use the change in energy splitting of
$|\Uparrow\downarrow\rangle$ and $|\Downarrow\uparrow\rangle$ PL
peaks as the measure of the Overhauser shift $E_{OHS}$}
\end{figure}

Nonzero average nuclear spin polarization along the external
magnetic field affects QD exciton energies via the hyperfine
interaction. This effect is demonstrated in Fig. \ref{fig:BNDet}
where PL spectra of the InGaAs QD A1 measured at $P_{exc}$=10~nW
in external magnetic field $B_z$=7~T are shown. The two spectra
were measured at $\sigma^+$ and $\sigma^-$ circularly polarized
optical excitation leading to different magnitudes of nuclear spin
polarization on the dot (see discussion in Sec.
\ref{subsec:DNPInGaAs}). Nuclear spin polarization shifts the
energies of the exciton PL lines\cite{HoleNucIso,GammonPRL}: the
sign of the shift is determined by the electron spin projection
($\uparrow$ or $\downarrow$) of that exciton.

In order to quantify the magnitude of the nuclear spin
polarization we use the changes of the energy splitting between
$|\Uparrow\downarrow\rangle$ and $|\Downarrow\uparrow\rangle$
excitons: $\Delta
E_{|\Uparrow\downarrow\rangle,|\Downarrow\uparrow\rangle}=E_{|\Uparrow\downarrow\rangle}-E_{|\Downarrow\uparrow\rangle}$.
Since the splitting $\Delta
E_{|\Uparrow\downarrow\rangle,|\Downarrow\uparrow\rangle}$ depends
on both magnetic field and nuclear spin polarization we determine
the Overhauser shift as a difference
\begin{eqnarray}
\label{eq:EOHSDef} E_{OHS}=-(\Delta
E_{|\Uparrow\downarrow\rangle,|\Downarrow\uparrow\rangle}-\Delta
E^{B_N=0}_{|\Uparrow\downarrow\rangle,|\Downarrow\uparrow\rangle}),
\end{eqnarray} where $\Delta
E^{B_N=0}_{|\Uparrow\downarrow\rangle,|\Downarrow\uparrow\rangle}$
is the splitting of the bright exciton doublet corresponding to
zero nuclear spin polarization. $\Delta
E^{B_N=0}_{|\Uparrow\downarrow\rangle,|\Downarrow\uparrow\rangle}$
is measured by either keeping the sample in the dark allowing
nuclear spins to relax\cite{InPDyn,InGaAsNucDyn}, or using
resonant radio-frequency excitation that depolarizes nuclear spins
\cite{QNMRArxiv,Bulutay}. Note the sign convention used:
$E_{OHS}>0$ corresponds to positive shift (increase in energy) of
the excitons with electron in a spin-up state $\uparrow$. The
magnitude of nuclear spin polarization can also be characterized
by the effective nuclear magnetic field $B_N=E_{OHS}/(g_e\mu_B)$.
By definition $B_N$ explicitly depends on the value of electron
g-factor $g_e$. We also note that in this work we neglect the
effect of the hole-nuclear spin interaction, since its net
contribution to $E_{OHS}$ is less than 5\% in the studied
structures \cite{HoleNucIso,FallahiHoleNuc}.

The total Overhauser shift $E_{OHS}$ is a sum of the Overhauser
shifts produced by different isotopes constituting the dot. The
contributions of different isotopes can in principle be separated
using nuclear magnetic resonance (NMR) techniques
\cite{QNMRArxiv}. However the phenomena described in this work are
not directly dependent on the way $E_{OHS}$ is distributed between
isotopes and thus we characterize nuclear spin polarization degree
using total Overhauser shift $E_{OHS}$.

\subsection{\label{subsec:DNPInGaAs}Dynamic nuclear polarization at ultra-low optical powers in InGaAs quantum dots}

\begin{figure}
\includegraphics{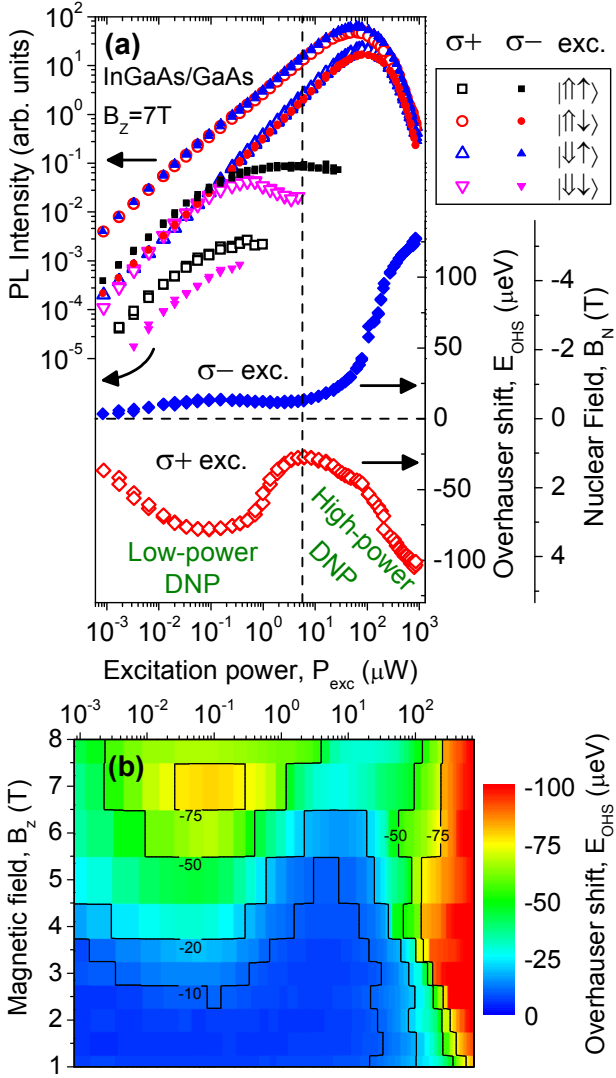}
\caption{\label{fig:PowDepInGaAs} (a) Results of power dependence
measurements on InGaAs neutral quantum dot QD A2 at $B_z=7$~T
under $\sigma^+$ (open symbols) and $\sigma^-$ (solid symbols)
optical pumping. PL intensities of all bright and dark exciton
states are shown at the top of the graph (left scale). Overhauser
shift $E_{OHS}=-(\Delta
E_{|\Uparrow\downarrow\rangle,|\Downarrow\uparrow\rangle}-\Delta
E^{B_N=0}_{|\Uparrow\downarrow\rangle,|\Downarrow\uparrow\rangle})$
is shown at the bottom of the graph with diamonds (right scale).
Additional scale on the right shows effective nuclear field $B_N$.
Vertical dashed line at $P_{exc}\sim5$~$\mu$W shows an approximate
boundary between two distinct nuclear spin pumping mechanisms:
low-power DNP via second order recombination of dark excitons and
high-power DNP due to spin transfer from spin polarized
excitons/electrons (see explanation in the text). (b) Contour-plot
of the power dependence of $E_{OHS}$ on $P_{exc}$ and $B_z$
measured for QD A2 under $\sigma^+$ optical pumping. Low-power DNP
is observed in a range of large magnetic fields with the maximum
$|E_{OHS}|$ achieved at $B_z=7$~T and $P_{exc}\approx100$~nW.}
\end{figure}

We now turn to the analysis of the mechanisms of the optically
induced dynamic nuclear polarization in the studied quantum dots.
For that we perform a series of power-dependent measurements on a
set of different quantum dots at different magnetic fields $B_z$.
In each measurement optical excitation power $P_{exc}$ is stepped
from high to low values in a wide range of more than six orders of
magnitude. After changing $P_{exc}$ the power is kept at this
level for $\sim$5~sec before PL spectrum is taken which is
sufficient to achieve the steady-state nuclear polarization and
eliminate any transient effects. From these spectra PL intensities
of excitons as well as the Overhauser shift $E_{OHS}$ are deduced
as a function of $P_{exc}$. The results of such an experiment done
on InGaAs QD A2 at $B_z=7$~T are presented in Fig.
\ref{fig:PowDepInGaAs}(a) for $\sigma^+$ polarized optical pumping
(open symbols) and $\sigma^-$ pumping (solid symbols).

PL intensities of all four exciton transitions are shown in the
top part of Fig. \ref{fig:PowDepInGaAs}(a) (left scale). The
intensities of bright excitons saturate at a power of
$\sim$80~$\mu$W, while dark excitons saturate at much lower power
$<5~\mu$W due to their significantly smaller oscillator strengths.

The power dependence of nuclear polarization in the same
measurement is shown in the bottom part of Fig.
\ref{fig:PowDepInGaAs}(a) (right scale). Two distinct regimes can
be observed. High-power DNP ($P_{exc}\gtrsim5~\mu$W) is
characterized by monotonic power dependence and direct
correspondence between the helicity of the light and the direction
of the resulting nuclear field. Such pattern in the DNP power
dependence is well studied
\cite{Skiba,InPX0,GaAsDiffusion,BrackerSSTReview}. In this regime
$\sigma^+$ ($\sigma^-$) optical pumping results in negative
(positive) Overhauser shift $E_{OHS}$ which exceeds $-$100~$\mu$eV
($+$120~$\mu$eV), corresponding to effective nuclear field in
excess of $+$4~T ($-$5~T). However these large values of $E_{OHS}$
are achieved only at very large pumping powers
$P_{exc}\sim1000~\mu$W for which exciton luminescence is saturated
and suppressed (Fig. \ref{fig:PowDepInGaAs}). Thus high-power DNP
can not be ascribed to ground state excitons and is likely to be a
result of nuclear spin polarization transfer from delocalized spin
polarized electrons in the wetting layer or highly excited QD
states \cite{InPX0}.

\begin{figure}
\includegraphics{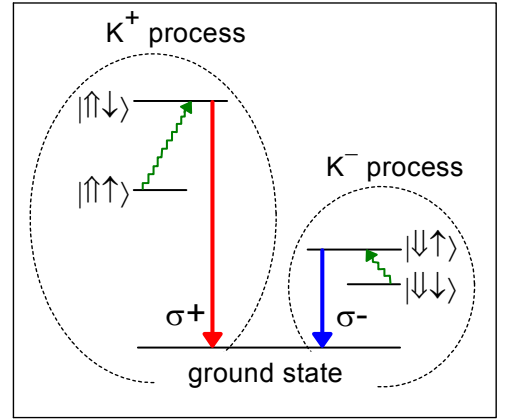}
\caption{\label{fig:NRGLevels} Energy level diagram of a neutral
exciton in an InGaAs/GaAs QD at high magnetic field $B_z\sim7$~T.
Vertical arrows show circularly polarized optical transitions due
to recombination of the bright $|\Uparrow\downarrow\rangle$ and
$|\Downarrow\uparrow\rangle$ excitons. Zigzag arrows show
transitions between dark ($|\Uparrow\uparrow\rangle$ and
$|\Downarrow\downarrow\rangle$) and bright excitons induced by
electron-nuclear hyperfine interaction. Two second-order processes
$K^+$ and $K^-$ are highlighted: each process starts from a QD
populated by a dark exciton ($|\Uparrow\uparrow\rangle$ or
$|\Downarrow\downarrow\rangle$ respectively) and changes total
nuclear spin polarization on the dot by +1 or -1 respectively (see
detailed explanation in Sec. \ref{subsec:DNPInGaAs}).}
\end{figure}

A significantly different nontrivial pattern is observed in the
low-power DNP regime ($P_{exc}\lesssim5~\mu$W). At $\sigma^+$
pumping Overhauser shift depends non-monotonically on the
excitation power with minimum $E_{OHS}\approx-80~\mu$eV observed
at a very low power of $P_{exc}\approx100$~nW. We note that with
the high-power DNP the same magnitude of $E_{OHS}$ can only be
achieved for excitation powers at least 3000 times higher.

Significant nuclear spin polarization observed at low excitation
powers can be explained by nuclear spin pumping via second-order
recombination of the dark excitons. This mechanism is similar to
that observed in InP/GaInP dots as reported in our previous work
(Ref. \cite{InPX0}). Figure \ref{fig:NRGLevels} gives a schematic
explanation. DNP takes place via transfer of spin polarization
from electrons to nuclei which requires electron to flip its spin.
In neutral excitons this requires conversion from a bright to a
dark state or vice versa. The electron spin-flips going in
opposite directions lead to increase or decrease of the total
nuclear spin polarization. However dark states have significantly
longer lifetimes making the processes starting from dark states
dominant.

There are two such processes\cite{KorenevSelfPol,InPX0}. The first
one denoted $K^+$ in Fig. \ref{fig:NRGLevels} is possible if QD is
occupied with $|\Uparrow\uparrow\rangle$ dark exciton: electron
can make a virtual flip converting exciton into a
$|\Uparrow\downarrow\rangle$ bright state and simultaneously
increasing total nuclear spin polarization by $+$1 (zigzag line in
the left half of Fig. \ref{fig:NRGLevels}). At the second stage
the virtual $|\Uparrow\downarrow\rangle$ exciton recombines
emitting $\sigma^+$ polarized photon. The other process ($K^-$)
starts from the $|\Downarrow\downarrow\rangle$ dark exciton and
goes via virtual $|\Downarrow\uparrow\rangle$ state (zigzag line
in the right half of Fig. \ref{fig:NRGLevels}). As a result of
this process $\sigma^-$ photon is emitted and the total nuclear
spin polarization is changed by $-$1. The $K^+$ and $K^-$ induce
nuclear spin polarization of opposite signs, thus the sign of the
overall nuclear spin pumping is determined by the process with
larger pumping rate.

The nuclear spin pumping rate of each of these second order
processes is proportional to two quantities \cite{InPX0}: (i)
population probability of the initial dark state, largely
determined by laser circular polarization (optically excited holes
rapidly lose their polarization during energy relaxation, while
electrons maintain their spin resulting in optical orientation of
dark excitons evidenced by dependence of PL intensities on
excitation helicity as seen in Fig. \ref{fig:PowDepInGaAs}(a)).
(ii) the probability of the virtual electron-nuclear flip-flop,
which in turn is inversly proportional to the square of the
splitting between the initial and intermediate states
[$\propto\Delta
E_{|\Uparrow\downarrow\rangle,|\Uparrow\uparrow\rangle}^{-2}$ for
process $K^+$ and $\propto\Delta
E_{|\Downarrow\downarrow\rangle,|\Downarrow\uparrow\rangle}^{-2}$
for process $K^-$].

At high magnetic field [$B_z\sim7$~T as in Fig.
\ref{fig:PowDepInGaAs}(a)] neutral exciton spectrum is asymmetric:
$\Delta
E_{|\Downarrow\downarrow\rangle,|\Downarrow\uparrow\rangle}<\Delta
E_{|\Uparrow\downarrow\rangle,|\Uparrow\uparrow\rangle}$ (for QD
A2 at $B_z=7$~T we have $\Delta
E_{|\Downarrow\downarrow\rangle,|\Downarrow\uparrow\rangle}^2/\Delta
E_{|\Uparrow\downarrow\rangle,|\Uparrow\uparrow\rangle}^2\sim1/5$).
As a result the efficiency of the $K^-$ process is enhanced,
leading to asymmetry in low-power DNP. Furthermore at $\sigma^+$
optical pumping $|\Downarrow\downarrow\rangle$ exciton is more
populated than $|\Uparrow\uparrow\rangle$ (this can be inferred
from PL intensities shown in Fig. \ref{fig:PowDepInGaAs}(a): in
the limit of low powers PL intensity of the
$|\Downarrow\downarrow\rangle$ exciton is a factor of $\sim$10
larger than that of $|\Uparrow\uparrow\rangle$). As a result $K^-$
process dominates over $K^+$ and leads to large negative
$E_{OHS}\sim-80~\mu$eV observed at low excitation power
$P_{exc}\sim100$~nW. By contrast at $\sigma^-$ pumping the
$|\Uparrow\uparrow\rangle$ is more populated favoring the $K^+$
process (PL intensity of the $|\Uparrow\uparrow\rangle$ exciton is
a factor of $\sim$70 larger compared to
$|\Downarrow\downarrow\rangle$). This however, is partially
compensated by the large splitting $\Delta
E_{|\Uparrow\downarrow\rangle,|\Uparrow\uparrow\rangle}$ of the
initial and intermediate states involved in $K^+$ process. As a
result of this partial compensation the overall Overhauser shift
does not exceed $E_{OHS}\approx+15~\mu$eV for $\sigma^-$
excitation.

In order to examine further the mechanism of DNP in neutral InGaAs
quantum dots we have performed a series of power-dependent
measurements at different magnetic fields $B_z$. The results for
QD A2 are shown in a contour plot in Fig.
\ref{fig:PowDepInGaAs}(b) where $E_{OHS}$ is plotted as a function
of $P_{exc}$ and $B_z$. It can be seen that $E_{OHS}$ induced at
low-powers ($P_{exc}\sim$100~nW) increases significantly with
magnetic field and reaches its maximum at $B_z\approx7$~T. Such
observation is in agreement with our interpretation that low-power
DNP is dominated by the second order recombination of the
$|\Downarrow\downarrow\rangle$ exciton: as expected the $\Delta
E_{|\Downarrow\downarrow\rangle,|\Downarrow\uparrow\rangle}$
splitting reduces with magnetic field [Fig. \ref{fig:BDep}(a)]
significantly enhancing the efficiency of the $K^-$ process. By
contrast the maximum $|E_{OHS}|$ achieved in the high-power DNP
($P_{exc}>$100~$\mu$W) is only weakly dependent on magnetic field
once it exceeds $B_z\gtrsim2$~T.

Thus we conclude that the low-power DNP observed for InGaAs/GaAs
dots (Fig. \ref{fig:PowDepInGaAs}) has the same origin as
low-power DNP previously observed in InP/GaInP QDs and the model
developed in Ref. \cite{InPX0} can be applied to InGaAs dots.
There are, however, some differences in DNP induced by dark
excitons in InP and InGaAs dots that are worth pointing out:

(i) In InGaAs dots $K^-$ is the most efficient process resulting
in large negative Overhauser shift $E_{OHS}$. By contrast in InP
dots DNP via dark excitons induces positive $E_{OHS}$. This is due
to the difference in electron and hole g-factors resulting in a
different energy level structure: in contrast to InGaAs in InP
dots the $\Delta
E_{|\Uparrow\downarrow\rangle,|\Uparrow\uparrow\rangle}$ splitting
is smaller than $\Delta
E_{|\Downarrow\downarrow\rangle,|\Downarrow\uparrow\rangle}$
resulting in higher efficiency of the $K^+$ process in InP.

(ii) In InP dots light-polarization-independent nuclear spin
pumping is observed: $E_{OHS}$ is positive for any degree of
circular polarization of light (including pure $\sigma^+$ and
$\sigma^-$), whereas the sign of low-power DNP in InGaAs dots is
controlled by the light helicity as described above. Such
difference can be explained by much higher fidelity of optical
orientation of the electron spin in InGaAs dots compared to InP:
in InGaAs dots PL intensity of the dark states changes by a factor
of $\sim$20 when excitation is changed between $\sigma^+$ and
$\sigma^-$ (Fig. \ref{fig:PowDepInGaAs}). By contrast for InP dots
this factor does not exceed $\sim$3, and the balance between $K^+$
and $K^-$ is controlled by the energy splitting of the initial and
intermediate states rather than by the population of the initial
states.

(iii) The low-power DNP process is generally more efficient in
InGaAs dots. The maximum absolute value of the Overhauser shift is
found to vary in the range $E_{OHS}\sim-80\div-50~\mu$eV in
different quantum dots in the same InGaAs sample (see Appendix
\ref{App:PDepAddit}) while for InP we previously observed maximum
$E_{OHS}\sim20\div30~\mu$eV. Such difference can be explained
qualitatively if we assume significantly longer dark exciton
lifetime in MBE-grown InGaAs dots compared to lower-quality
MOVPE-grown InP dots, where dark state population decays due to
charge fluctuations resulting from a high level of impurities.
Further insight in this direction can be achieved by experiments
on QD-structures where dark exciton lifetime can be manipulated
directly and independently (e.g. using electric bias in
Schottky-diode structures or using microwave excitation).

\subsection{\label{subsec:DNPGaAs}GaAs quantum dots: Modification of the low-power DNP due to electron-hole exchange interaction}

\begin{figure}
\includegraphics{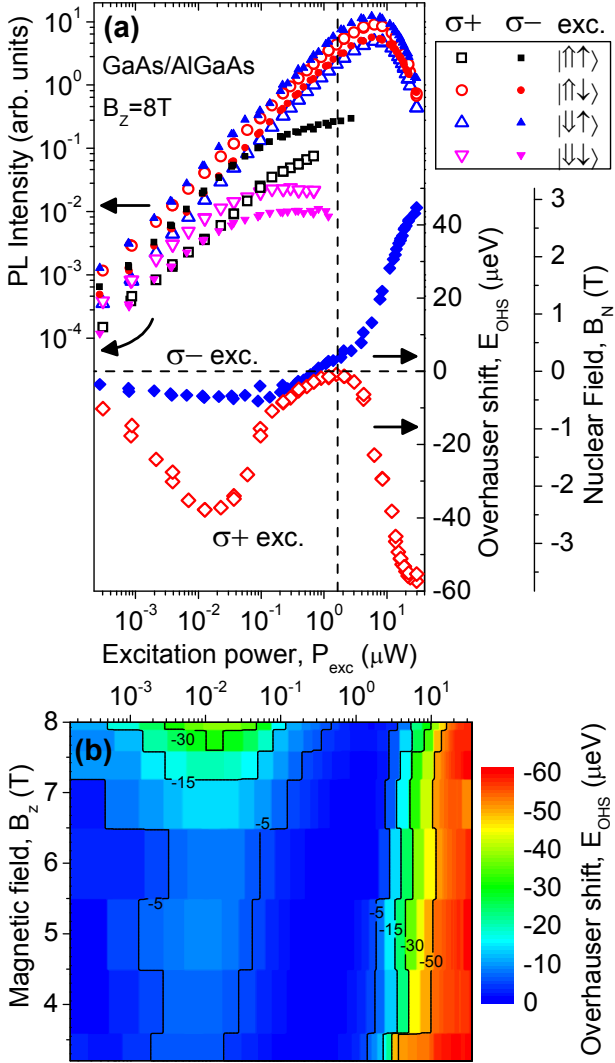}
\caption{\label{fig:PowDepGaAs} (a) Results of power dependence
measurements on GaAs neutral quantum dot QD B1 at $B_z=8$~T under
$\sigma^+$ (open symbols) and $\sigma^-$ (solid symbols) optical
pumping. PL intensities of all bright and dark exciton states are
shown at the top of the graph (left scale). Overhauser shift
$E_{OHS}$ is shown at the bottom of the graph with diamonds (right
scale). Additional scale on the right shows effective nuclear
field $B_N$. (b) Contour plot of the power dependence of $E_{OHS}$
on $P_{exc}$ and $B_z$ measured for QD B1 under $\sigma^+$ optical
pumping.}
\end{figure}

We have also performed power dependence measurements on
GaAs/AlGaAs QDs at different magnetic fields. The results for QD
B1 are shown in Fig. \ref{fig:PowDepGaAs}. Similar to InGaAs dots
two regimes in DNP are clearly observed. The low-power DNP
mechanism reaches its peak efficiency at $B_{z}\approx8$~T,
$P_{exc}\approx10$~nW resulting in $E_{OHS}\approx40~\mu$eV at
$\sigma^+$ polarized optical pumping.

There is a significant difference in the results for GaAs dots
compared to InGaAs: it follows from Fig. \ref{fig:PowDepGaAs} that
low-power DNP via dark excitons leads to negative $E_{OHS}$ in
GaAs dots despite the fact that high magnetic field reduces the
$\Delta E_{|\Uparrow\downarrow\rangle,|\Uparrow\uparrow\rangle}$
splitting [see Fig. \ref{fig:BDep}(b)] which should in principle
enhance the $K^+$ process efficiency and lead to a net positive
$E_{OHS}$. Furthermore, it can be seen from Fig.
\ref{fig:PowDepGaAs}(a) that low-power DNP is more efficient at
$\sigma^+$ pumping which also results in increased population of
the $|\Downarrow\downarrow\rangle$ dark state. This allows us to
conclude that low-power DNP in QD B1 is dominated by the $K^-$
process despite its large splitting $\Delta
E_{|\Downarrow\downarrow\rangle,|\Downarrow\uparrow\rangle}$ of
the initial and intermediate states.

These seemingly contradicting results can be explained if we take
into account the role of anisotropic electron-hole exchange
(Coulomb) interaction which has no significant effect in case of
InGaAs dots. Quantum dot symmetry reduction can lead to mixing of
bright and dark states: such mixing manifests itself as repulsion
and anticrossing of the bright and dark lines in PL spectra
measured in magnetic field along the growth axis
\cite{Bayer,ChekhovichSPD}. In the studied GaAs dots anticrossing
of the $|\Uparrow\uparrow\rangle$ and $|\Uparrow\downarrow\rangle$
states takes place at $B_z>8$~T and thus could not be observed
directly. However dark-bright mixing manifests itself as increased
saturated PL intensity of the $|\Uparrow\uparrow\rangle$ state
compared to $|\Downarrow\downarrow\rangle$ at $B_z=8$~T (see Fig.
\ref{fig:PowDepGaAs}(a)).

As we have shown previously for InP dots the dark-bright mixing
resulting from low-symmetry exchange interaction suppresses
low-power DNP near the anticrossing point. Such suppression
results from reduction of the dark exciton lifetime and reduction
of the electron spin projections of the exciton eigenstates
\cite{InPX0}. In InP and InGaAs dots dark-bright mixing results in
reduction of $|E_{OHS}|$ induced by the dominant second-order
process ($K^-$ in InGaAs dots and $K^+$ in InP) in the small
vicinity of the anticrossing point: in InGaAs QD A2 $|E_{OHS}|$
decreases when magnetic field is increased from 7~T to 8~T despite
reduction of the $\Delta
E_{|\Downarrow\downarrow\rangle,|\Downarrow\uparrow\rangle}$
splitting (see Figs. \ref{fig:BDep}(a) and \ref{fig:PowDepInGaAs}
(b)). By contrast in GaAs dot B1 the effect of the exchange
interaction is much stronger and it completely suppresses the
$K^+$ process making $K^-$ dominant and resulting in negative
$E_{OHS}$ in the whole range of magnetic fields used (see Fig.
\ref{fig:PowDepGaAs}(b)). Thus the increase of $|E_{OHS}|$ at
$B_z=8$~T observed in QD B1 is due to suppression of the process
with the small dark-bright splitting ($K^+$) which makes the
process with larger dark-bright splitting ($K^-$) dominant.

Dark-bright mixing induced by exchange interaction depends
strongly on the dot symmetry and thus can change significantly
from dot to dot in the same sample \cite{Bayer}. As a result the
magnitude of the low-power DNP also changes appreciably which has
been observed in the measurements performed on several GaAs dots
(see Appendix \ref{App:PDepAddit}). In particular we find that for
some GaAs dots $E_{OHS}$ induced in the low-power regime changes
sign and becomes positive implying that dark-bright mixing in such
dots is reduced compared to QD B1. On the other hand measurements
on several InGaAs dots reveal very close values of $E_{OHS}$ of
the same sign supporting our interpretation that low-symmetry
exchange interaction is small in the studied InGaAs dots.

\section{\label{sec:Concl}Conclusions}

We have studied experimentally dynamic nuclear polarization under
ultra-low power non-resonant optical excitation in individual
InGaAs/GaAs and GaAs/AlGaAs quantum dots. We have demonstrated
that nuclear spin pumping via second-order recombination of dark
neutral excitons previously observed in InP/GaInP dots
\cite{InPX0} has a general nature and is found in InGaAs/GaAs and
GaAs/AlGaAs dots. In InGaAs dots low-power optical pumping is
found to lead to large Overhauser shifts up to $\sim80~\mu$eV. In
GaAs dots low-power DNP is systematically found to be less
efficient than in InGaAs dots: we attribute this to the
dark-bright exciton mixing stemming from the low-symmetry
electron-hole exchange interaction. We find that the varying
magnitude of the dark-bright mixing in different individual
GaAs/AlGaAs dots leads to variations in the magnitude of the
Overhauser shift including the change of its sign.

\section{Acknowledgments}

This work has been supported by the EPSRC Programme Grants
EP/G001642/1 and EP/J007544/1 and the Royal Society. J. Puebla
gratefully thanks CONACYT-Mexico Doctoral Scholarship.

\appendix
\section{\label{App:PDepAddit}Additional experimental results on dynamic nuclear polarization
in neutral QDs at ultra-low power optical pumping}

\begin{figure*}
\includegraphics{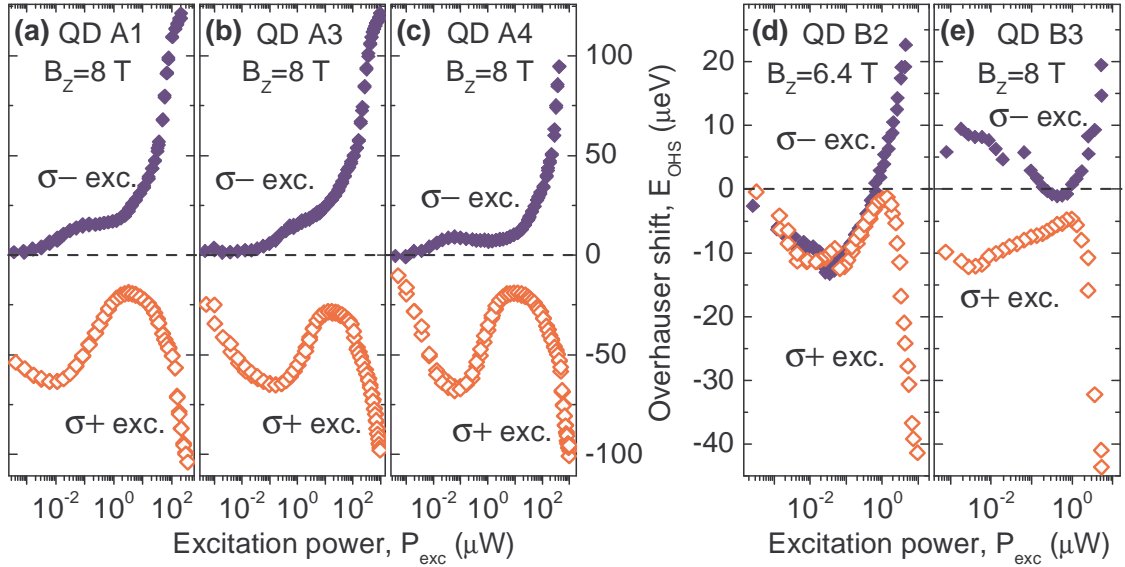}
\caption{\label{fig:PDepAddit} Results of power dependence
measurements on several individual neutral QDs from the same
InGaAs and GaAs samples as reported in Figs.
\ref{fig:PowDepInGaAs}, \ref{fig:PowDepGaAs}. (a-c) show the data
for InGaAs QDs A1, A3 and A4 measured at $B_z=8$~T [compare to
Fig. \ref{fig:PowDepInGaAs}(a)]. (d,e) show the data for GaAs QDs
B2 and B3 measured at $B_z=6.4$~T and $B_z=8$~T respectively
[compare to Fig. \ref{fig:PowDepGaAs}(a)].}
\end{figure*}

In order to verify systematically the nature of the low-power DNP
we have performed power dependent measurements on a set of few
individual quantum dots in the same InGaAs/GaAs and GaAs/AlGaAs
samples. Fig. \ref{fig:PDepAddit} shows the results for three more
InGaAs dots A1, A3 and A4 (a-c) and two more GaAs dots B2 and B3
(d,e).

For InGaAs dots we find results very similar to dot A2 (Fig.
\ref{fig:PowDepInGaAs}): low power DNP results in peak
$E_{OHS}\approx-70~\mu$eV observed for $\sigma^+$ optical pumping
at $B_z=8$~T. By contrast, we find different behavior for
different GaAs dots. In QD B2 we find negative $E_{OHS}$ for both
$\sigma^+$ and $\sigma^-$ optical pumping suggesting that similar
to the case of QD B1 dark-bright exciton mixing strongly
suppresses the $K^+$ process. On the other hand in QD B3 we find
that the sign of the low-power DNP is controlled by the helicity
of the optical excitation: under $\sigma^-$ pumping which enhances
$|\Uparrow\uparrow\rangle$ initial population the $K^+$ becomes
dominant leading to $E_{OHS}>0$. The recovery of the $K^+$ process
observed in QD B3 can be explained by the reduction of the
anisotropic $e-h$ exchange interaction in this dot.


\end{document}